\documentclass[a4paper]{article}
\usepackage{graphicx}
\usepackage{subcaption}
\usepackage{booktabs}
\usepackage{listings}
\usepackage[htt]{hyphenat}
\usepackage[hidelinks]{hyperref}

\usepackage{array}
\newcommand{\PreserveBackslash}[1]{\let\temp=\\#1\let\\=\temp}
\newcolumntype{C}[1]{>{\PreserveBackslash\centering}p{#1}}
\newcolumntype{R}[1]{>{\PreserveBackslash\raggedleft}p{#1}}
\newcolumntype{L}[1]{>{\PreserveBackslash\raggedright}p{#1}}

\hypersetup{pdftitle={A New Model for Testing IPv6 Fragment Handling},pdfauthor={Edoardo Di Paolo, Enrico Bassetti and Angelo Spognardi}}

\begin{document}

\title{A New Model for Testing IPv6 Fragment Handling}

\author{Edoardo Di Paolo, Enrico Bassetti, Angelo Spognardi}
\date{\{dipaolo,spognardi,bassetti\}@di.uniroma1.it}

\maketitle

\begin{abstract}
Since the origins of the Internet, various vulnerabilities exploiting the IP fragmentation process have plagued IPv4 protocol, many leading to a wide range of attacks. IPv6 modified the handling of fragmentations and introduced a specific extension header, not solving the related problems, as proved by extensive literature. One of the primary sources of problems has been the overlapping fragments, which result in unexpected or malicious packets when reassembled. To overcome the problem related to fragmentation, the authors of RFC 5722 decided that IPv6 hosts MUST silently drop overlapping fragments.

Since then, several studies have proposed methodologies to check if IPv6 hosts accept overlapping fragments and are still vulnerable to related attacks. However, some of the above methodologies have not been proven complete or need to be more accurate.
In this paper we propose a novel model to check IPv6 fragmentation handling specifically suited for the reassembling strategies of modern operating systems. Previous models, indeed, considered OS reassembly policy as byte-based. However, nowadays, reassembly policies are fragment-based, making previous models inadequate. Our model leverages the commutative property of the checksum, simplifying the whole assessing process.
Starting with this new model, we were able to better evaluate the RFC-5722 and RFC-9099 compliance of modern operating systems against fragmentation handling. Our results suggest that IPv6 fragmentation can still be considered a threat and that more effort is needed to solve related security issues.

\end{abstract}

\section{Introduction}\label{sec:introduction}

Internet standards allow the use of fragmentation when a router has to transmit an IP packet larger than the next link's \textit{Maximum Transmission Unit} (MTU), i.e., the maximum number of bytes that the link can transmit in a single IP packet.
The fragmentation process consists of dividing the packet into smaller units, called fragments, so that the resulting pieces can pass through a link with a smaller MTU than the original packet size.

Initial IPv4 specifications, RFC~791\cite{RFC0791}, describes a reassembly algorithm that allows new fragments to overwrite any overlapped portions of previously-received fragments~\cite{RFC1858}.
Unfortunately, this algorithm enabled bypassing filtering solutions and resulted in the operating system adopting different policies to reassemble fragments~\cite{novak2005target}.
Over the years, various vulnerabilities that exploit the fragmentation process have been discovered, mainly using overlapping fragments, exposing the Internet to several types of attacks: \textit{Denial of Service} (DoS), \textit{Traffic Modification}, \textit{Traffic Interception}, \textit{Intrusion Detection Systems} (IDS)/\textit{Intrusion Prevention Systems} (IPS) \textit{evasion}, \textit{Firewall evasion}~\cite{gilad_herzberg_2013,shankar_paxson}.

IPv6 brought about significant changes to handling fragmentation compared to its predecessor, IPv4. It introduced a specific extension header and aimed to address the shortcomings of IPv4 fragmentation. RFC~5722~\cite{RFC5722} tackles the fragmentation problem by explicitly forbidding overlapping fragments. However, despite these efforts, extensive literature and previous studies have demonstrated that IPv6 fragmentation still poses security risks. In particular, it has been shown that many operating systems are not entirely RFC~5722 compliant, accepting some sequences of IPv6 overlapping fragments, being exposed to several forms of detection evasion~\cite{attacking} and traffic hijacking~\cite{gilad_herzberg_2013}.


Numerous studies have been conducted to assess the vulnerability of IPv6 hosts to overlapping fragments and related attacks. However, some of the existing methodologies have not been proven to be complete or accurate enough. Some others, like the Shankar and Paxson model, were proposed in the past, but they are obsolete due to recent changes in the reassembly strategies, as we will demonstrate in this work. Therefore, we propose a novel model specifically designed to evaluate the handling of IPv6 fragmentation, taking into account the reassembling strategies employed by modern operating systems.


To prove the usefulness of our model, we thoughtfully tested it over widely used operating systems. We also compared the results achieved using the Shankar and Paxson model over the same targets. As shown later, our model was able to capture the non-compliance of all operating systems that we tested, whereas the Shankar and Paxson model indicates full compliance on IPv6 fragmentation.

Additionally, to demonstrate that IPv6 fragmentation has still to be considered a real threat, we implemented a \textit{Traffic Modification} attack. The attack requires the ability to predict the IP identification number (IP-id): IP-id prediction has been considered quite critical since a long time ago, and some successful attempts are present in literature~\cite{10.5555/1051646,zalewski1,zalewski2,pmtuharmful,idwild}. In the attack, we take advantage of the partial or non-existing compliance of RFC~5722 to alter the legitimate traffic between two hosts, again exploiting the use of overlapping fragments~\cite{attacking}.

Thus, we show that vulnerabilities inherent to IPv6 fragmentation persist. Despite numerous recommendations, attacks on IPv6 fragmentation remain feasible, necessitating more effort to eliminate all flaws in implementations.

The paper is structured as follows: the next section provides a brief background on IP fragmentation and past work on the topic. Section~\ref{sec:ip-fragm-handl} introduces two well-known models for testing IPv6 fragmentation issues, and discusses their limitations. Section~\ref{sec:sec-new-model} reports our experiments performed to evaluate the RFC~5722 compliance of modern operating systems. Section~\ref{sec:rfc9099} reports our findings on RFC~9099 compliance. Section~\ref{sec:manip-attacks-with} report our experiment results for the Traffic Modification attacks. Finally, Section~\ref{sec:conclusions} summarizes the contributions of our work and provides some further comments to help fix the IPv6 fragmentation flaws.

\section{Background and Related Works}\label{sec:background-about-ip}

In this section, we briefly introduce some details about IP fragmentation that provide the background for the experimental section. Then, we report a quick survey about the main contribution related to IP fragmentation vulnerabilities, focusing on IPv6.

\subsection{IP Fragmentation in Internet}\label{sec:ip-fragm-ipv6}

An essential property of an Internet link is the number of bytes it can transmit in a single IP packet, namely the \textit{Maximum Transmission Unit} (MTU). The MTU may differ between different networking technologies. IPv4 requires every link to support a minimum MTU of 576 bytes, as recommended by RFC 791~\cite{RFC0791}, while IPv6 requires every link to support an MTU of 1280 bytes or greater (RFC 2460~\cite{RFC2460}). When an endpoint has to transmit an IP packet greater than the next link MTU, IP calls for fragmentation, which is the process of separating a packet into units (fragments) smaller than the link MTU. The receiving host performs fragment reassembly to pass the complete (re-assembled) IP packet up to its protocol stack.

The fragmentation process is handled differently in IPv4 and IPv6. In IPv4, an IP packet can be fragmented by the source node or intermediate routers along the path between the source and the destination. However, intermediate routers may avoid fragmenting IPv4 packets by dropping the packet and forcing the Path MTU by the source host. In IPv6, only \textit{end-to-end fragmentation} is supported; intermediate routers cannot create fragments. To discover the best MTU size, both IPv4 and IPv6 leverage on the \textit{Path MTU Discovery}, provided by \textit{Internet Control Message Protocol} version 4 (ICMP) or 6 (ICMPv6).

In order to reassemble all the fragments related to the same packet, IP protocol uses some information present in the header, namely: the \textit{identification} field (shared among all the fragments of the same packet), the \textit{fragment offset} (that specifies which is the starting fragment position in the original packet) and the \textit{More Fragments} flag, set to 1 for all fragments except for the last one. IPv4 and IPv6 differ mainly on the identification field length (16-bit long in the former, 32-bit in the latter), and that IPv6 uses a specific extension header to hold the above information.

While not predominant, statistics show that IP fragmentation is still used in the Internet and, notably, for security protocols like IKE~\cite{kaufman2010internet} or DNSSEC~\cite{arends2005dns}, that typically rely on large UDP packets for cryptographic material exchange~\cite{gilad_herzberg_2013}.

\subsection{Related Works}\label{sec:related-works}

IP Fragmentation has been exploited to make many different types of attacks, as anticipated by Mogul and Kantarjiev back in 1987~\cite{10.1145/55482.55524}. Most of them allow realizing denial of services~\cite{kaufman_perlman_sommerfeld_2003} or IDS/firewall evasion~\cite{shankar_paxson} or operating system fingerprinting~\cite{attacking}. Besides those based on IP fragmentation, many other attacks rely on the possibility of predicting the IP identification field of the victim\cite{10.5555/1051646}. For this reason, there has been a flurry of works focusing on the feasibility of predicting IP-id field~\cite{10.1145/1533057.1533063,qian_mao_2012,qian_mao_xie_2012,10.5555/1251398.1251411}. While most studies about IP fragmentation are related to IPv4, only a few specifically focus on IPv6. As it will be further explored in Section~\ref{sec:ip-fragm-handl}, IPv6, among the other vulnerabilities~\cite{ullrich2014ipv6} mitigations, has been revised to specifically fix the IP fragmentation issues, firstly with RFCs 3128~\cite{miller2001tiny}, 5722~\cite{RFC5722}, 6946~\cite{gont2013atomic} and, later, with an updated version of the IPv6 specification, namely RFC 8200~\cite{RFC8200}. Moreover, RFC 9099 ~\cite{rfc9099} discusses extension headers with a special focus on fragments stating that, if not handled properly, they could lead to bypassing stateless filtering.

In~\cite{kaufman_perlman_sommerfeld_2003}, the authors exploit the IP fragmentation to prevent legitimate IKE handshakes from completing. The main idea is to flood one of the endpoints with fragments to consume all the memory resources dedicated to the fragment buffers, realizing a fragment cache overflow. The overflow prevents legitimate hosts from completing the IKE handshake because the reassembling becomes impossible. Another type of attack has been described in RFC 4963~\cite{RFC4963}, consisting of a fragment misassociation. The idea is that the attacker can poison the fragment cache of a host, sending some spoofed fragments so that when the fragments of the victim reach the poisoned host, they will be misassociated and, consequently, maliciously altered. 

The most influential work for our research has been done by Gilad and Herzberg~\cite{gilad_herzberg_2013}. They present a DoS attack inspired by a traffic injection technique based on IP fragmentation, proposed by Zalewski\footnote{Michal Zalewski, \textit{A new TCP/IP blind data injection technique?}, \url{https://seclists.org/bugtraq/2003/Dec/161}} and appeared in the seclist mailing list in 2003. The idea is to inject the second fragment of TCP connections since the IP-id was highly predictable. Following this intuition, the authors in~\cite{gilad_herzberg_2013} propose performing a DoS attack against a communicating host, targeting the NAT-ing host behind which the destination endpoint resides. This type of attack allows the authors to realize a very effective DoS attack, causing more than 94\% of packet loss without leveraging any fragment cache overflow. 

Another pivotal work that inspired this paper has been done by Atlasis~\cite{attacking}. In his paper, the author performs an exhaustive battery of tests to verify the effective behavior of several operating systems when overlapping IP fragments are present. In particular, the experiments verified the different reassembly strategies and how those can be exploited to perform various evasion attacks. The methodology used in Atlasis' paper for understanding the different reassembly policies is the driving factor for realizing our Traffic Modification attack, as detailed in Section~\ref{sec:manip-attacks-with}. Moreover, Atlasis in ~\cite{atlasisevasion} shows how high-end commercial IDPS devices could be evaded by the use of IPv6 extension headers.
\section{IP Fragmentation Handling in the Wild}\label{sec:ip-fragm-handl}

In this section, we evaluate RFC 5722 compliance of different operating systems. We introduce two established methodologies adopted in the literature to evaluate the IP fragmentation reassembly strategies that we discovered to be obsolete. Then, we propose a new methodology, based on the presented ones, and we discuss the results we obtained by testing widely used operating system.

\subsection{Shankar and Paxson Model}\label{sec:shank-pax}

\begin{figure}[htb]
    \centering
    \begin{minipage}{.55\textwidth}
        \centering
        \includegraphics[width=\textwidth]{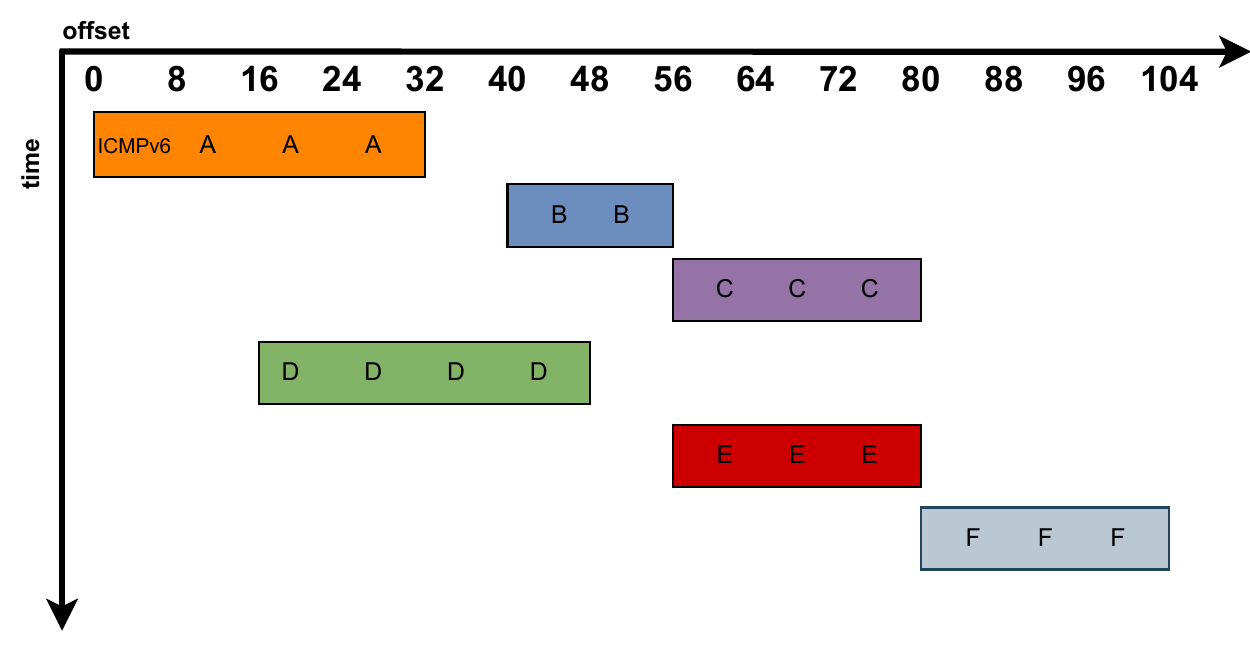}
        \caption{Shankar and Paxson model.}
        \label{fig:shankar}
    \end{minipage}%
    \begin{minipage}{.45\textwidth}
        \centering
        \includegraphics[width=\textwidth]{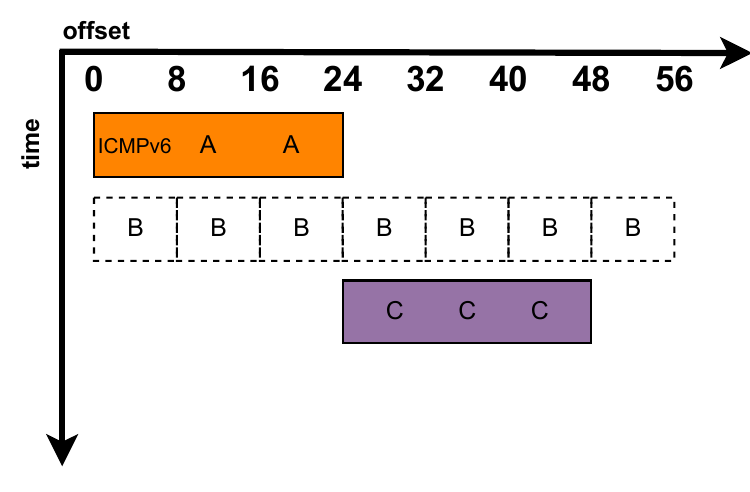}
        \caption{Three Fragments Model}
        \label{fig:3frag}
    \end{minipage}
\end{figure}

The first methodology we consider is the one we call \textit{Shankar and Paxson model}~\cite{shankar_paxson}. In their paper, the authors introduce a model consisting of six fragments of different lengths and offsets, as shown in Figure~\ref{fig:shankar}, creating a diversified combination of fragment overlap and overwrite. In the figure, each fragment is represented by a block labeled with a character (e.g., `A', `B', `C'), and the payload of each fragment is a sequence of bytes encoding the corresponding character. The vertical axis marked as ``time'' represents the temporal succession of the transmitted fragments. For example, the first fragment in Figure~\ref{fig:shankar} has offset 0 in the final (reassembled) payload, a length of 32 bytes, and contains the ICMPv6 header plus 24 `A's.

For each two adjacent fragments, X and Y, the Shankar and Paxson model guarantees that there is~\cite{shankar_paxson}:

\begin{itemize}
\item At least one fragment (X) wholly overlapped by a subsequent fragment (Y) with identical offset and length;
\item At least one fragment (X) partially overlapped by a subsequent fragment (Y) with an offset greater than fragment X;
\item At least one fragment (X) partially overlapped by a subsequent fragment (Y) with an offset smaller than fragment X.
\end{itemize}

By using six fragments, five different fragment reassembled sequences were found~\cite{novak2005target}:
\textit{BSD}, that favors an original fragment with an offset smaller or equal to a subsequent fragment;
\textit{BSD-right}, that favors a subsequent fragment when the original fragment has an offset smaller or equal to the subsequent one;
\textit{Linux}, that favors an original fragment with an offset that is smaller than a subsequent fragment;
\textit{First}, that favors the original fragment with a given offset;
\textit{Last}, that favors the subsequent fragment with a given offset.

In our experiments we discovered that modern operating systems don't use parts of a fragment: they assemble fragments by using or discarding them entirely, as discussed in Section~\ref{sec:overlapping-today}. Due to this new behavior, the Shankar and Paxson model is no more suitable for IPv6 fragment overlapping tests as the reassembly phase may never finish in some occasion (more on this in Section~\ref{sec:overlapping-today}).

\subsection{Three Fragments Model} \label{sec:3-frag}

Besides the Shankar and Paxson model, another methodology named \textit{``3-fragments model''} was proposed by Atlasis~\cite{attacking}. They used this methodology to evaluate a host behavior with the fragmentation overlapping. Their model is based on several tests in which only three fragments are exchanged, and only the header and payload of the second fragment change, as shown in Figure~\ref{fig:3frag}.

The model is defined by these three fragments:

\begin{itemize}
    \item The first fragment has always offset 0, and \textit{More Fragments} flag (\textit{M-flag} from now on) set to 1. It consists of an ICMPv6 Echo Request (8 bytes) with 16 bytes of payload for a total length of 24 bytes;
    \item The second fragment has variable length and offset, and within the different tests, it also varies in the value of the \textit{M-flag};
    \item The third fragment has always offset 24, \textit{M-flag} always set to 0, and a length of 24 bytes, carrying part of the payload. 
\end{itemize}

The model comprises 11 different combinations of length and offsets for the second fragment while varying the value of the \textit{M-flag} for the second fragment and reversing the sending order of the three fragments (from 1 to 3 and from 3 to 1). This shuffling leads to a total of 44 tests.

While this test was successfully used to investigate the IP fragmentation reassembly~\cite{attacking}, the model is now obsolete because it assumes that IPv6 endpoints may reassemble the packet using a fragment partially. As discussed in Section~\ref{sec:overlapping-today}, modern operating systems assemble fragments by using or discarding them entirely.

\section{A New Model for Testing IPv6 Fragment Handling}\label{sec:sec-new-model}

This section proposes a new model to check RFC~5722 compliance on IPv6 fragmentation. We discuss how operating systems handle IPv6 fragments nowadays, and then we present our proposal for a model for testing overlapping fragments based on the Shankar and Paxson model. Finally, we discuss the results obtained from the different experiments performed.

\subsection{Overlapping Fragments Today}\label{sec:overlapping-today}

\begin{figure}[tb]
    \centering
    \includegraphics[width=\textwidth]{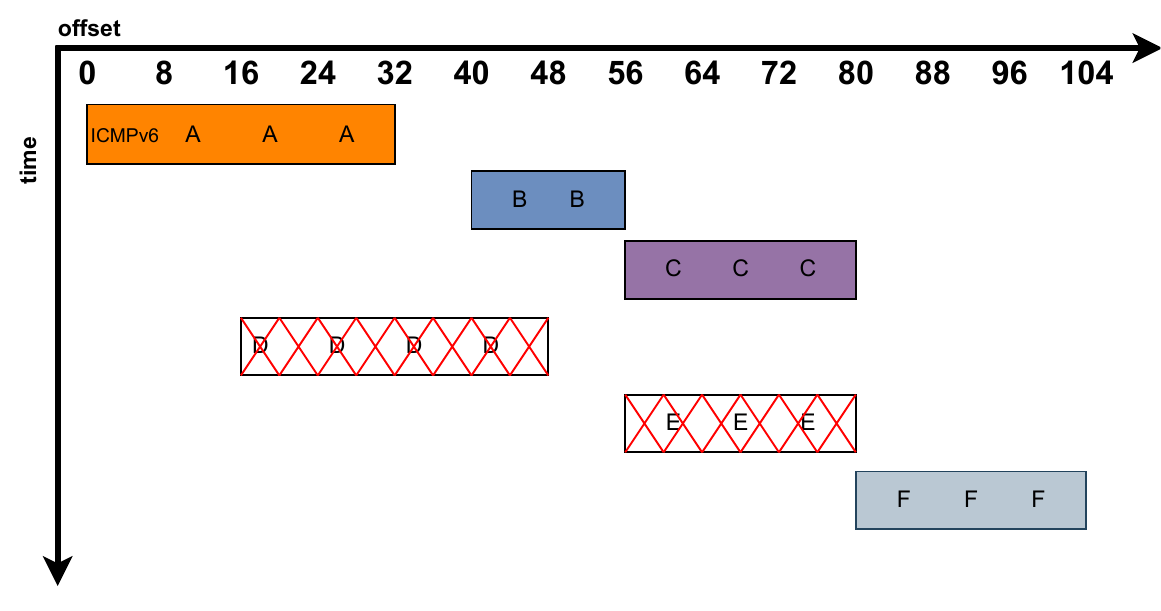}
    \caption{Overlapping fragments discarded by operating systems may create hole and prevent a correct reassembly. In this example, the operating system decided to discard ``D" and ``E" fragments due to its reassembly policy, leaving a hole between ``A" and ``B".}
    \label{fig:model_discard_overlapping}
\end{figure}

Previously proposed models for testing overlapping fragments proved to be obsolete in our experiments. We noted that all operating systems in Table~\ref{tab:os_tested} discard entire overlapping fragments. Which fragment is discarded depends on the operating system policy, which may include the arrival time or offset position. Figure~\ref{fig:model_discard_overlapping} shows an example of this problem with the Shankar and Paxson model when the operating system drops overlapping fragments that arrived late.
Since fragments ``D" and ``E" overlap with ``A", ``B" and ``C", they are discarded by the operating systems, thus producing a ``hole'' between fragments ``A'' and ``B".

The gap between ``A" and ``B" (caused by dropping ``D") does not allow the machine to reassemble the packet correctly and reply to the ``ICMPv6 Echo Request'', since it is missing information between offsets 32 and 40. The machine waits for a predetermined time (\texttt{ip6frag\_time}, which in some systems is set to 30 seconds by default) and then deletes the fragments received up to that time from memory. Thus, these machines seem compliant with the RFC~5722, as there is no way to know externally whether the packet has been discarded because of the gap or because they dropped the packet and all fragments (which is the action required by the RFC).

\subsection{A New Model for Testing IPv6 Fragment Handling}

\begin{figure}[tb]
    \centering
    \includegraphics[width=\textwidth]{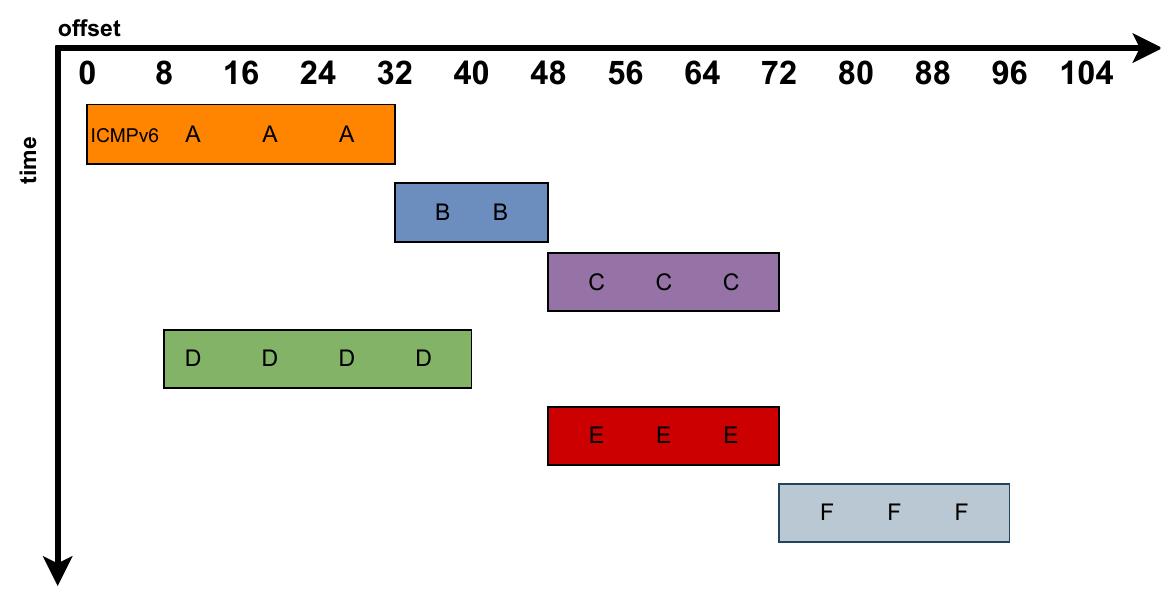}
    \caption{The new model we are proposing. It is based on the Shankar and Paxson model but it has different offsets to better fit current reassembly policies.}
    \label{fig:our_model}
\end{figure}

Our approach is based on the well-known Shankar and Paxson model, which provides a comprehensive framework for analyzing the reassembly process of fragmented packets. However, we modify the original model by reducing all fragment offsets by one, resulting in an offset reduction of 8 bytes. Also, although we display the model using the same time sequence as the Shankar and Paxson model, our model is meant to be tested by shuffling the sequence. In other words, multiple tests should be done, and the fragments should have different arrival times on each test but the same offsets and content. The model is shown in \autoref{fig:our_model}.

The primary motivation behind the offset mutation is to have some combinations where the packet reassembly is done by using or discarding entire fragments, as modern operating systems do. Previous models could not expose issues in the fragmentation handling due to the problem discussed in Section~\ref{sec:overlapping-today}.

It is important to note that the ICMPv6 header is excluded from this overlap. We excluded the ``next header" in the payload as no operating systems allow overwriting it in any combination of fragments.

A significant contribution of our modified model lies in identifying two combinations of fragments, AAABBCCCFFF and AAABBEEEFFF, that do not create any holes during reassembly. These fragment combinations are particularly interesting, as they present scenarios where fragments can be reassembled even when partial overlapping fragments are dropped.

To avoid issues with different checksums created by different ways of reassembling fragments, we also re-defined the payload of the fragments as shown in Table~\ref{tab:payload-def} and presented in Section~\ref{sec:test-checksum}. By doing so, we have that any combination of fragments in our model has the same checksum.

\begin{figure}[tb]
    \centering
    \includegraphics[width=\textwidth]{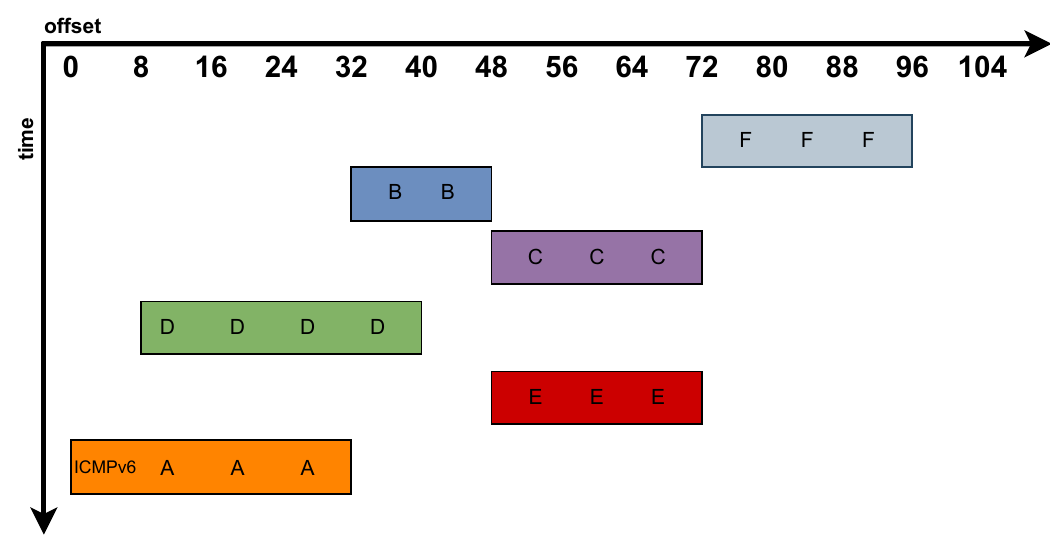}
    \caption{One of the permutation is when the first fragment is the latest one to arrive. This particular arrangement might result in a different reassembled packet.}
    \label{fig:first_last}
\end{figure}

\subsection{Model Validation and Results}\label{sec:result-discussion}

We performed a series of experiments to assess the usefulness of our model and to catch anomalies in the reassembly procedure in operating systems. For some tests, we also compared our model with Shankar and Paxson, demonstrating that our model is able to capture the non-compliance where the Shankar and Paxson model suggests that the operating system is RFC~5722-compliant. We also tested all possible permutations of the arrival time of the fragments (720 permutations). Although some may be superfluous due to previous tests, we checked the complete set of permutations to create a dataset for further analysis. Overall, the total number of tests in our dataset is 2226\footnote{The complete list of tests can be found in the \href{https://github.com/netsecuritylab/ipv6-fragmentation}{GitHub repository}.} for each operating system listed in Table~\ref{tab:os_tested}.

\setlength{\tabcolsep}{9pt}
\begin{table}[tb]
\centering
\caption{Operating systems, versions tested and Vagrant box version.}
\begin{tabular}{ccc}
\toprule
\textbf{Operating System} & \textbf{Kernel Version} & \textbf{Vagrant box version} \\
\midrule
GNU/Linux Arch & 6.3.2-arch1-1 & 3.23.0522 \\
GNU/Linux Debian 11 & 5.10.0-22-amd64 & 11.20230501.1 \\
GNU/Linux Ubuntu 23.04 & 6.2.0-20-generic & 20230524.0.0 \\
OpenBSD & 6.9 GENERIC.MP\#473 & 4.2.16 \\
FreeBSD & 13.1-RELEASE & 4.2.16 \\
Microsoft Windows 10 & 10.0.19041.2965 & 2202.0.2305 \\
Microsoft Windows 11 & 10.0.22621.1702 & 2202.0.2305 \\
\bottomrule
\end{tabular}
\label{tab:os_tested}
\end{table}

All tests were performed in a small laboratory created by Vagrant by Hashicorp, an open-source software that helps and simplifies the management of reproducible virtual machine environments. The lab network contains one attacker, a Debian 11 virtual machine with our fuzzer, and one victim, which rotated between all operating systems listed in Table~\ref{tab:os_tested}. The victim and the attacker are directly connected, as in a LAN, to avoid side effects by other network elements. We performed tests multiple times to reduce the number of errors due to some transient situation in the victim.

We can summarize the tests in:
\begin{enumerate}
    \item Single ICMPv6 packet fragmented using multiple permutations of fragments;
    \item Single ICMPv6 packet fragmented using multiple permutations, but fragments are sent multiple times;
    \item Multiple ICMPv6 packets fragmented using multiple permutations.
\end{enumerate}

In the first test, a single packet is fragmented and sent to the victim for each different permutation in our model. In the second test, the same occurs, but all frames are sent again (in the same order) 4 more times, to simulate a network retransmission or a malicious act by the attacker. In the third test, for each different permutation, five different ICMPv6 packets are created and sent. Note that, while both the second and the third tests are sending five packets, in the second test all fragments have the same fragment ID (as they belong to the same packet), while in the third test fragments are grouped by the fragment ID, which is different for each packet.

\begin{table}[tb]
\centering
\caption{ICMPv6 Echo replies received while testing the Paxson and Shakar model and our model. A RFC-compliant host should never reply to these tests.}
\label{tab:model-experiments-results}
\begin{tabular}{cC{2cm}C{0.6cm}C{0.6cm}C{0.6cm}}
\hline
\textbf{}                       & \textbf{Shankar and Paxson} & \multicolumn{3}{c}{\textbf{Proposed model}} \\ 
\hline
\textbf{OS / Test} &       & \textbf{1}    & \textbf{2}   & \textbf{3}   \\
\hline
GNU/Linux Arch                  &    0         & 35            & 37           & 1634         \\
GNU/Linux Debian                &   0      & 35            & 37           & 1634         \\
GNU/Linux Ubuntu               &    0    & 35            & 37           & 1634         \\
OpenBSD                         &   0      & 0            & 24           & 373        \\
FreeBSD                         &   0      & 0            & 20            & 1609          \\
Windows 10                      &   0      & 0            & 20            & 1800 \\
Windows 11                      &   0      & 0            & 20            & 1800        \\
\hline
\end{tabular}
\end{table}

Table~\ref{tab:model-experiments-results} shows a summary of the results we obtained by running the proposed model in the virtual environment. These results not only demonstrate that all listed operating systems are violating the RFC~5722 by replying to overlapping fragments, but also that the way they handle overlapping fragments (the ``reassembly policy") may pose some risk, as in the case when different reassembly policies are used between an IDS/IPS/firewall and the victim. Different reassembly policies may result in a different reassembled packet: an attacker can exploit this difference to provide a different packet to the IDS/IPS/firewall than the one reassembled by the victim, bypassing the IDS/IPS/firewall protection.

We performed the same set of tests using the Paxson and Shakar model. As shown in Table~\ref{tab:model-experiments-results}, the Paxson and Shakar model ended with no replies for all tests, which may indicate RFC~5722 compliance. However, we know that ICMPv6 Echo Replies are missing not because of discarding overlapping fragment (RFC requirement), but due to missing payload pieces in the reassembly created by the new policy of fragment reassembly (as explained in Section~\ref{sec:overlapping-today}).

\subsection{On IPv6 Checksum and Overlapping Fragments}\label{sec:test-checksum}

IPv6 header does not contain a checksum~\cite{RFC8200}, shifting the duty of checking the integrity of the transmission to the upper layers. The choice was made to speed up the packet forwarding: IPv6 intermediate devices, like routers, do not check the integrity of the datagram (except for security systems like IDS). Upper layers protocols, like ICMPv6, UDP, and TCP, may require a checksum, which is computed and verified by the transmission endpoints.

When the checksum is required (e.g., ICMPv6), the first fragment contains the checksum of the entire packet inside the upper layer header. Overlapping fragments might create a situation where the final checksum of the packet is incorrect since hosts can have different reassembly policies~\cite{attacking}. This, in turn, might cause the victim to discard the packet and not reply to our ICMPv6 ping tests, invalidating the results presented until now.

To rule out that packets are dropped because of the checksum mismatch, we re-defined the payload of the fragments as shown in Table~\ref{tab:payload-def}. This new definition exploits the commutative property of the checksum: by definition, the checksum contains the sum of 2 bytes pair of the entire IPv6 packet ~\cite{rfc4443}. Given that the sum has the commutative property, any combination of fragments in our model has the same checksum.

\begin{table}[tb]
\centering
\caption{New payload definition. This payload exploits the checksum's commutative property to avoid re-assembly errors. The ``odd'' version is also used in tests with one packet, whereas both ``odd'' and ``even'' are used in tests with multiple packets.}
\begin{tabular}{ccc}
\toprule
\textbf{Shankar and Paxson} & \textbf{Odd/single packet} & \textbf{Even}     \\ 
\midrule
A                     & 11223344   & 44113322 \\
B                     & 11332244   & 44331122 \\
C                     & 22113344   & 44332211 \\
D                     & 22331144   & 11224433 \\
E                     & 33112244   & 11334422 \\
F                     & 33221144   & 22114433 \\
\bottomrule
\end{tabular}
\label{tab:payload-def}
\end{table}

\subsection{Denial of Service due to RFC 5722 Compliance}\label{sec:dos-attack-with}

The RFC 5722 requires dropping packets with overlapping fragments. This rule creates a vulnerable surface for a Denial-of-Service attack by a malicious third host, the ``attacker". For the attacker to exploit the vulnerability, the transmission between two parties should use fragments, and the attacker should be able to predict the IP-id field and spoof the source IP address.

The attack strategy involves the attacker sending a spoofed fragment to the target host (the transmission receiver) before it can reassemble all the fragments received from the victim (the transmission source). By using the same source address as the victim and the same IP-id value, the attacker creates an overlap during the fragment reassembly phase, causing the entire packet to be discarded and preventing the victim from receiving a reply.

This Denial-of-Service vulnerability exposed by the RFC is exploitable only in certain conditions. Also, it is the desired effect as fragment handling exposes higher security issues. So, in our opinion, it does not present a real threat due to the practical difficulties of satisfying prerequisites and the low gain of such attacks.

\section{RFC 9099 Compliance}\label{sec:rfc9099}

This section presents the study on RFC 9099 \cite{rfc9099} compliance, which mainly focuses on IPv6 Extension Headers. We briefly introduce the requirements stated in RFC~9099~\cite{rfc9099}. Then, we describe the experiments we performed to verify the compliance of operating systems.

\subsection{IPv6 Fragment Headers and RFC~9099}

IPv6 extension headers are additional data structures that can be included in the IPv6 packet header to provide extra functionality or options beyond what is defined in the IPv6 protocol. These headers are placed between the IPv6 header and the upper-layer protocol data, allowing for features such as fragmentation, authentication, encryption, routing, and mobility.

IPv6 uses a specific extension header named ``Fragment Header" to handle a fragmented packet where the packet is too big for the Maximum Transfer Unit of underlying links~\cite{RFC8200}. While not strictly required, RFC~9099~\cite{rfc9099} suggest that security devices in the middle of the transmission (such as firewall, IDS, IPS) and the destination endpoint should drop first fragments that do not contain the entire IPv6 header chain (which include the transport-layer header). The reason for this requirement is to avoid issues when dealing with stateless filtering~\cite{RFC8200}.

Some IPv6 extension headers can be expressed multiple times in a single datagram; others do not. Also, these headers are linked together in a chain, and the RFC~8200~\cite{RFC8200} suggests an order for headers to optimize the computation and avoid processing issues. RFC~9099 strongly recommends that the correct order and the maximum number of repetitions of extension headers are enforced in endpoints and in any security device in the path. Non-conforming packets should be dropped\cite{rfc9099}.

In the context of fragmented IPv6 packets, a malicious actor may try to send additional headers in different fragments, or they might try to overwrite the upper layer header in the payload from the first fragment using a subsequent overlapping fragment. In both cases, the packet must be discarded.

\subsection{Fragment Headers Experiments and Results}

We designed our experiments around the two requirements from RFC~9099: the first experiment checks the requirement of having all extension headers in the first fragment, and the second experiment is on overlapping the upper layer header using an overlapping fragment.

In the first experiment (\autoref{table:rfc9099_first}), we send a packet divided into three fragments: the first contains the ICMPv6 Header and no payload, the second contains the payload, and the last contains a Destination Options header. These fragments are not overlapping, but no response is expected as the first fragment does not contain the complete IPv6 headers chain~\cite{rfc9099}.

In the second experiment (\autoref{table:rfc9099_second}), we send a packet divided into three fragments: the first contains no headers and no payload, the second contains the ICMPv6 Echo Request header (offset for this fragment is zero, as the first fragment), and the last contains a payload. The first and the second fragments do not overlap because the first fragment is empty. Nevertheless, no response is expected as the first fragment does not contain the complete IPv6 headers chain~\cite{rfc9099}.

\begin{table}[tb]
\centering
\caption{First experiment for RFC~9099 compliance. No response is expected: the first fragment should be dropped because it does not contain all the extension headers.}
\setlength{\tabcolsep}{6pt}
\begin{tabular}{ccccc}
\toprule
\textbf{\#} & \textbf{Type}                       & \textbf{Offset} & \textbf{More Fragment} & \textbf{Payload} \\ \midrule
1           & ICMPv6 Echo Request                 & 0               & 1                      & /                \\
2           & Fragment                            & 1               & 1                      & AAAAAAAA         \\
3           & IPv6 Fragment \& Dest. Options & 2               & 0                      & BBBBBBBB         \\ \bottomrule
\end{tabular}
\label{table:rfc9099_first}
\end{table}

\begin{table}[tb]
\caption{Second experiment for RFC~9099 compliance. No response is expected: the first fragment should be dropped because it does not contain the full IPv6 headers chain.}
\centering
\begin{tabular}{ccccc}
\toprule
\textbf{\#} & \textbf{Type} & \textbf{Offset} & \textbf{More Fragment} & \textbf{Payload} \\
\midrule
1 & IPv6 Fragment & 0 & 1 & / \\
2 & ICMPv6 Echo Request & 0 & 1 & / \\
3 & IPv6 Fragment & 1 & 0 & BBBBBBBB \\
\bottomrule
\end{tabular}
\label{table:rfc9099_second}
\end{table}

We tested a subset of the operating system in Table~\ref{tab:os_tested}, namely Debian 11, FreeBSD, and OpenBSD.

In the first experiment, we received an ICMPv6 Echo Reply from all the operating systems, even if the first fragment did not have the destination options header. The correct response was to discard the first fragment or the entire packet.

In the second experiment, all operating systems recognized the malformed packet; however, only FreeBSD and OpenBSD silently dropped the received packet. Debian, instead, the victim responds to the attacker with an ICMPv6 packet with code 3, which stands for ``IPv6 First Fragment has incomplete IPv6 Header Chain".

The results of the first experiment show that the systems tested, at present, are not compliant with RFC~9099 since they are not dropping the first fragments or packets when extension headers are spread through fragments. Also, since test 1 contains the ``Destination Options" extension header after the Fragment Header (while the RFC~8200 recommends the opposite), those systems are not dropping non-conforming packets, while RFC~9099 suggests discarding them.

Moreover, it is possible to recognize an operating system using the fragmented packet as described in \autoref{table:rfc9099_second} (second experiment): the different behavior between operating systems could lead an attacker to perform fingerprinting against victims. While RFCs~9099 and 5722 do not specify whether a host should silently discard these packets, we believe a silent discard is the safest option, and an RFC should mandate it.

\section{Modification Attacks with Overlapping Fragments}\label{sec:manip-attacks-with}

This type of attack has been firstly performed, in a different context, by Gilad et al.~\cite{gilad_herzberg_2013}: the attacker aims to modify the content of the communication between the victim and a legitimate host, namely host X. We leverage the RFC~5722 non-compliance: since the victim accepts overlapping fragments, the attacker can change with another fragment the bytes sent by the host X.

In the Modification Attack, the attacker must send one or more fragments and ensure that when these fragments are reassembled with the legitimate ones, the result is a correct packet (not malformed) and that the final packet payload is the desired one. For this reason, to perform a successful modification attack, the attacker has to calculate the correct checksum to avoid the discard of a packet because of a checksum mismatch (Section~\ref{sec:test-checksum}).

We demonstrate that the modification attack is possible by altering the content of a syslog UDP transmission from host X to the victim.

\subsection{Implementation of Modification Attack with Scapy}
\label{sec:modification-Scapy}

We present the implementation of the Modification Attack in a specific scenario where both host X and the victim are running the \texttt{rsyslog}, an open-source software for managing logs and forwarding log messages between machines. The host X is configured to send syslog messages via UDP to the victim.

This attack requires the malicious actor to know (or guess) the IP identification field value, the payload of the IPv6 packet (in this case, the log message) host X sends to the victim, and requires spoofing the host X address. The IP ID requirement can be easily satisfied: guessing the IP identification value has already been demonstrated in literature by Salutari et al.~\cite{idwild}. The packet payload, instead, should be predictable (for example, a TLS Client-Hello) or known in advance. Spoofing is still a problem, especially in local area networks~\cite{idwild}.

We will alter the log line regarding a successful SSH authentication in this case. In particular, we will try to alter the fragment containing the attacker's IP address. The original line is \texttt{Jun 1 20:47:08 git sshd[88459]: Accepted publickey for git from 10.10.10.100 port 49240 ssh2: ED25519 \\SHA256:vNTXCU7b6C6mqvcaH7j1/uRC5unllTpG5kCtd01xxoc}

The host X sends the log line in three fragments:

\begin{enumerate}
    \item the first fragment with the UDP header, the syslog severity and facility code, and the first 51 bytes of the message:  \\ \texttt{<43> Jun 1 20:47:08 git sshd[88459]: Accepted publickey };
    \item the second fragment with 56 bytes of the message: \\ \texttt{for git from 10.10.10.100 port 49240 ssh2: ED25519 SHA25};
    \item the third fragment with the rest of the log line: \\ \texttt{6:vNTXCU7b6C6mqvcaH7j1/uRC5unllTpG5kCtd01xxoc};
\end{enumerate}

To successfully perform the attack, the malicious actor should send a spoofed second fragment with a different payload before or after the second fragment from the host X to the victim (depending on the reassembly policy used by the victim). When the victim re-assembles the final packet, the new payload for the second fragment will be used.

However, a different payload will likely result in a different checksum: the victim will drop the packet as corrupted. To work around this problem, the attacker exploits the commutative property of the sum in the checksum by shuffling the payload. The shuffle should swap groups of two bytes as the checksum is calculated by a series of 16-bit sums.

Another technique for keeping the same checksum is to calculate the difference between the correct checksum and the checksum of the new final packet and then add this difference to the spoofed fragment payload. Thanks to the commutative property of the sum, the result will be the original checksum.

\subsection{Modification Attacks Results Discussion}

In our test environment, we used two GNU/Linux Debian machines (the same version listed in Table~\ref{tab:os_tested}). We tested this attack by sending the following payload in a spoofed second fragment to the victim just before the original second fragment: \texttt{0 0. Efromr 9225HAgi.12: 110fo55po10D240rt19t 0. 4sh S s}. This string is a permutation of the payload of the original second fragment, so the final payload (after the reassembly) will remain the same as the original one.

\begin{figure}[tb]
    \centering
    \includegraphics[width=\textwidth]{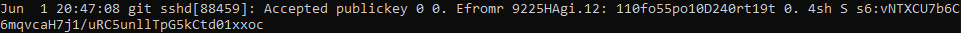}
    \caption{The string in the log file. The log line contains the attacker's modified payload.}
    \label{fig:syslog_attack_proof}
\end{figure}

Since the RFC~5722 states that datagrams with overlapping fragments must be silently discarded, we should not expect any log in the victim machine. However, as the victim is not RFC~5722 compliant, we found the modified payload in the log file, as shown in \autoref{fig:syslog_attack_proof}.

These attacks can be prevented mainly by implementing RFC~5722 recommendations of dropping the entire packets in the presence of overlapping fragments. IPSec may provide additional protections.


\section{Conclusions}\label{sec:conclusions}

In conclusion, this work addresses the ongoing issues associated with packet fragmentation in IPv6, explicitly focusing on the issue of overlapping fragments. Despite the requirement listed by different RFCs for hosts to drop overlapping fragments silently, our work indicates that the problem persists. Also, changes in the fragment reassembly policies by operating systems from byte-based to fragment-based made current models for testing IPv6 fragmentation issues (such as the Shankar and Paxson model) obsolete.

To address these issues, the authors propose a novel model that exploits the fragment-based strategy in modern operating systems when handling IPv6 fragmentation. By leveraging the commutative property of the checksum, the authors simplify the assessment process and propose a more accurate evaluation methodology.

Using this new model, the authors evaluate the compliance of modern operating systems with RFC-5722 and RFC-9099, which pertain to fragmentation handling in IPv6. The evaluation was performed both using ICMPv6 Echo Request/Reply, and by performing a real attack named ``Modification Attack'', where a fragmented transmission was altered.

The results of the evaluation reveal that IPv6 fragmentation remains a significant threat, and further efforts are required to address the related security issues. These findings underscore the need for ongoing research and development to enhance the security measures and mechanisms associated with IPv6 fragmentation.

Taking the necessary countermeasures to deal with fragmentation attacks and secure IPv6 would still be appropriate since adopting IPv6 is an irreversible and ever-growing process, especially with new technologies based on the Internet of Things.

We released the dataset and all scripts developed to run our experiments in a public GitHub repository at \\ \url{https://github.com/netsecuritylab/ipv6-fragmentation}.

\section*{Acknowledgments}

This work was partially supported by project ‘Prebunking: predicting and mitigating coordinated inauthentic behaviors in social media’ project, funded by Sapienza University of Rome; by the Italian Ministry of Defense PNRM project ``UNAVOX'', by project SERICS (PE00000014) under the MUR National Recovery and Resilience Plan funded by the European Union -- NextGenerationEU.

\bibliographystyle{abbrv}
\bibliography{bibliography}

\end{document}